# Effect of different precursors on CVD growth of molybdenum disulfide

Aditya Singh*, Monika Moun, and Rajendra Singh

*Department of Physics, Indian Institute of Technology Delhi, New Delhi 110016, India*

* adityasingh27993@gmail.com

**Abstract**

Control over thickness, size, and area of chemical vapor deposition (CVD) grown molybdenum disulfide ($MoS_2$) flakes is crucial for device application. Herein, we report a quantitative comparison of CVD synthesis of $MoS_2$ on $SiO_2$/Si substrate using three different precursors viz., molybdenum trioxide ($MoO_3$), ammonium heptamolybdate (AHM), and tellurium (Te). A three-step chemical reaction mechanism of evolution of $MoS_2$ from $MoO_3$ micro-crystals is proposed for $MoO_3$ precursor. Furthermore, a strategy based on growth temperature and ratio of amount of precursors is developed to systematically control thickness and area of $MoS_2$ flakes. Our findings show that for large-sized crystalline monolayer $MoS_2$ flakes, $MoO_3$ is a better choice than AHM and Te-assisted synthesis. Moreover, Te as growth promoter, can lower down growth temperature by ~250°C. This study can be further used to fabricate $MoS_2$ based high-performance electronic devices such as photodetectors, thin film transistors, and sensors.

## 1. Introduction

Two-dimensional (2D) materials, namely graphene, black phosphorus, hexagonal boron nitride, and transition metal dichalcogenides (TMDCs) have gained tremendous attention due to their unique optical, electrical, and mechanical properties[1,2]. Graphene being a zero bandgap 2D-material, has limited applications in opto-electronics devices[3]. Possible ideal materials for replacement of graphene in future are TMDCs, depicted by $TX_2$, where T and X are transition metal (Mo, W, Nb, Ti etc.) and chalcogen (S, Se, and Te), respectively e.g., $MoS_2$, $MoSe_2$, and $MoTe_2$ etc[1]. In most of the semiconducting TMDCs, a transition from an indirect bandgap to a direct bandgap has been observed as number of layers decrease due to the quantum confinement effect[2]. Molybdenum disulphide ($MoS_2$), a member of TMDCs family, has been widely explored due to tunability of bandgap (1.3 eV in bulk and 1.8 eV in monolayer)[4], high elastic modulus (~170 N/m)[5], high on/off ratio (~$10^8$)[6], high carrier mobility (40-480 $cm^2$/V-s)[7], and lack of short channel effect[7], which makes it appropriate for application in flexible devices[5,7], high performance field-effect-transistors (FETs)[7,8], photodetectors[9,10] and gas sensors[11] etc.

Inter-layer strong covalent and intra-layer weak van der Waals bonding makes peeling of $MoS_2$ layers easier[1,2]. Till now, various methods have been developed to grow layered $MoS_2$ such as mechanical exfoliation (scotch tape) [7,12], physical vapor deposition (sputtering and pulsed laser deposition)[13], one-step CVD (direct reaction of S and Mo source)[13], and two-step CVD (sulfurization of pre-deposited Mo film)[14,15] etc. Mechanical exfoliation method was employed to obtain high-quality layered $MoS_2$ films but this approach limits its industrial growth and commercially viable device applications because of low yield and small size $MoS_2$ films/flakes[16,17]. Furthermore, the hydrothermal method was also employed to synthesize $MoS_2$ but failed due to the same limitations as of exfoliation[1,18,19]. Highly crystalline homogeneous monolayer $MoS_2$ with high yield and low cost having excellent electrical and optical properties is desirable. So, special attention has been given to CVD as it has a greater possibility to grow large area high-quality uniform mono- and few-layer $MoS_2$ with low manufacturing costs[20–22].

CVD growth of $MoS_2$ could be achieved by different Mo source materials such as molybdenum trioxide ($MoO_3$) powder[19,23–25], Mo film[26], ammonium heptamolybdate[21,27], gas source of $Mo(CO)_6$[20,23,27], and $MoCl_5$[20,23]. Zhan *et al.*[28] reported the growth of $MoS_2$ film on silicon dioxide ($SiO_2$) substrate by sulfurization of Mo metal film in a two-step CVD process but this approach resulted in a lack of crystallinity and poor electrical properties of $MoS_2$ films[29]. In a single-step process, Mo metal film was replaced by Mo sources such as $MoO_3$ powder, AHM or $MoCl_5$. $MoS_2$ flakes grown by $MoCl_5$ have superior uniformity and good transport properties but $MoCl_5$ being toxic can make growth hazardous[29]. For $MoS_2$ growth on the $SiO_2$/Si substrate, $MoO_3$ and S precursors have been dominantly used because of preferential highly crystalline monolayer $MoS_2$ having quite good electrical and optical properties at the wafer scale[30]. Cain *et al.*[31] proposed the mechanism for the $MoS_2$ nucleation, growth, and grain boundary formation to study systematically, and prepare large area, single- and few-layered TMDCs films. Gong *et al.* [32] suggested that use of Te can reduce the growth temperature of $MoS_2$, and Tao *et al.*[21] developed low-cost CVD method to grow centimetre size highly crystalline $MoS_2$ films by ammonium molybdate as Mo source precursor.

Our present work focuses on the effect of different Mo precursors ($MoO_3$ and AHM) on the shape, size, and crystallinity of monolayer (1L), bilayer (2L), and multilayer (bulk) $MoS_2$ flakes on $SiO_2$/Si substrate. Also, we used Te as growth intermediate along with $MoO_3$ to reduce growth temperature by ~250°C. It was observed that choice of precursors influences the geometry, crystallinity, and surface morphology of as-grown $MoS_2$ flakes.

## 2. Experimental

Single zone atmospheric pressure mini CVD (APCVD) tube system (MTI Corporation, OTF-1200X S50-2F) was used for the growth of 1L- and few-layer MoS$_2$ using different precursors such as sulfur (S) powder (99.98%, Sigma-Aldrich, 414980), MoO$_3$ powder (99.97%, Sigma-Aldrich, 203815), AHM [(NH$_4$)$_6$Mo$_7$O$_{24}$· 4H$_2$O)] (99.98%, Sigma-Aldrich, 431346), and Te powder (99.997%, Sigma-Aldrich, 264865) on Si (001) n-type substrate covered with thermally grown 285 nm thick SiO$_2$ layer. The precursors and the substrates were placed in a 1.97 inch diameter and 17.7 inch long quartz tube, which was surrounded by the furnace (**Fig. 1a**). We did not perform any pre-growth treatment, such as seeded growth, use of molecular sieves or any step edges that may have adverse effects on electrical properties of flakes[21]. The substrates were cleaned with acetone followed by IPA and DI water, and placed face down on the top of boat shape quartz crucible (5.5 cm × 2 cm × 0.5 cm). Mo source was placed inside quartz boat under face down SiO$_2$/Si substrate at hottest zone (700-950°C) of the tube, and S powder was kept 16 cm away from MoO$_3$ at a low-temperature gradient of the quartz tube. After both precursors and substrates were loaded into the tube, ~60 ml/min of argon (Ar) gas (~99%, Sigma Gases), controlled by mass flow controller, was introduced into the tube to purge the system. After all these processes, the furnace was started at a ramp rate of 10°C/min. **Fig. 1b** represents the CVD temperature-time graph for the temperature of Mo source versus time. At 300°C, the temperature was kept constant for 5 min so that all contaminants, pre-occupied precursors, moisture, and dust can be swept out by Ar gas flow. Melting and vaporisation of S were started at 600°C and 900°C (temperature at the centre of the tube), respectively, and the temperature was kept constant at 900°C for 20 min. Vaporised S particles were carried by Ar gas over the substrate where it reacted with vaporised MoO$_3$ particles, and formed different shape few- and 1L-MoS$_2$ flakes. The system was cooled down to room temperature, and byproducts were carried by continuous Ar gas flow. **Fig. 1c** shows image of as-grown $MoS_2$ sample on $SiO_2$/Si substrate showing large area growth. Blue boxes represent regions of grown $MoS_2$ flakes while white boxes show $MoO_2$ and/or $MoOS_2$ grown regions.

Optical microscopy (OM) images of as-grown $MoS_2$ flakes of different shapes (triangles, truncated triangles, hexagons, bowtie, pine leaves, and stars) on $SiO_2$/Si substrate were captured by Nikon Eclipse LV100 microscope. Both Raman spectroscopy and photoluminescence measurements were carried out at room temperature by Renishaw-inVia-confocal-Raman microscope-6260 with 532 nm laser line and 1800 lines/mm grating on as-grown $MoS_2$ flakes to observe fundamental modes of $MoS_2$ ($E_{2g}^1$ and $A_{1g}$)[33] and their bandgaps, respectively. Layer numbers of crystalline $MoS_2$ flakes and its surface roughness were determined by atomic force microscopy (Bruker, Dimension Icon) in tapping mode with Si tip. To observe the surface morphology and elemental contrast, field emission scanning electron microscopy (FESEM) was carried out by FESEM-Zeiss microscope in back scattering mode.

## 3. Results and discussion

Uniform large size high quality $MoS_2$ flakes on $SiO_2$/Si substrate were grown by CVD using three different methods as illustrated below.

### 3.1. Growth of $MoS_2$ using S and $MoO_3$

$MoS_2$ thin layers were grown using $MoO_3$ and S powder by systematically optimising the growth parameters such as amount of precursors and growth temperature. Initially, the S content (20-500 mg) was optimized with the constant loading of 20 mg $MoO_3$. Considering the concept of mole, 1 mg of S (atomic weight ~32) contains ~$1.88 \times 10^{19}$ atoms while 1 mg of $MoO_3$ (atomic weight ~144) contains ~$4.18 \times 10^{18}$ molecules. At S=20 mg (S to $MoO_3$ particles ratio ~5:1), T= 700°C, and at vacuum (~5 torr), no proper deposition was obtained. Inside the quartz tube, S vapor particles are in the downstream flow, and atomically, they need to travel very large distance (16 cm) to reach over the $SiO_2$/Si substrate where they will react with $MoO_3$ vapor particles. Most of S vapor particles go unreacted and flushed out of the outlet. So, we need to increase the amount of S in order to increase the reaction possibility. When the S amount was doubled to 40 mg (S to $MoO_3$ particles ratio ~10:1) that resulted in formation of clusters and grains of oxy-particles of $MoO_3$ ($MoO_2$) uniformly all over the substrate as shown in **Fig. 2a**. Again when the amount of S was further increased to 80 mg (S to $MoO_3$ particles ratio ~18:1) and temperature to 800°C at atmospheric pressure, small-sized (~3 μm) oxy-phase of $MoS_2$ ($MoOS_2$) flakes were obtained (**Fig. 2b**). Insufficient supply of S caused the formation of $MoO_2$ or $MoOS_2$, and these oxy-phases acted as nucleation centres for $MoS_2$ growth[25]. When amount of $MoO_3$ was decreased from 20 mg to 15 mg (S to $MoO_3$ particles ratio ~30:1), large triangles (~20-80 μm) of 1L-$MoS_2$ on the large area (~3 mm × 4 mm) were observed (**Fig. 2c**). **Fig. 2c** shows the growth of a few 2L-$MoS_2$ triangular flakes (~2 μm) over 1L-$MoS_2$ flakes. Lateral growth of 1L-$MoS_2$ flakes over $SiO_2$/Si substrate was caused by lateral strong covalent bonding while growth of small-sized 2L-$MoS_2$ flakes over 1L-$MoS_2$ was caused by vertical weak van der Waals epitaxy[34]. Bright triangular flake at the bottom right of **Fig. 2c** shows a multilayer growth of $MoS_2$. So, at more S vapor particles than $MoO_3$ (S to $MoO_3$ particles ratio ~30:1), crystalline $MoS_2$ flakes can be easily grown. **Fig. 2d-i** represent FESEM images (in back-scattering mode) of as-grown $MoS_2$ flakes synthesised by S and $MoO_3$ precursors representing the evolution of $MoS_2$ from $MoO_3$ oxy-phase ($MoO_2$) to highly crystalline uniform large-sized triangular 1L-$MoS_2$ flakes. These FESEM images clearly show the optical and compositional contrast of flakes with respect to SiO$_2$/Si substrate. **Fig. 2d,e** represent FESEM images of MoO$_2$ and MoOS$_2$ flakes corresponding to OM images 2(a) and(b), respectively. **Fig. 2f** shows rectangular bulk MoS$_2$ micro-crystals deposited at the boundary of contact between substrate and quartz boat. **Fig. 2g** depicts polycrystalline continuous bulk MoS$_2$ growth along with few single crystalline MoS$_2$ triangular flakes, and **Fig. 2h** shows a few hexagonal flakes along with truncated triangles which may be considered as the basic building block of triangular MoS$_2$ flakes[19]. In **Fig. 2i**, MoS$_2$ triangular flakes are very sharp and have uniform surface morphology, depicting very high single crystallinity.

Sulfurization of MoO$_3$ shows the transformation of MoO$_3$ to MoS$_2$ involving three intermediate steps[15]. At first, MoO$_3$ powder thermally evaporates at ~700°C, reduced using S, and re-condense to obtain MoO$_2$ micro-crystals where each removed oxygen atom partakes in forming SO$_2$. A possible chemical reaction is:

$$2MoO_3 + S \rightarrow 2MoO_2 + SO_2 \uparrow \tag{1}$$

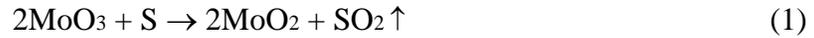

In the second step, MoO$_2$ micro crystals again sulfurize under appropriate conditions at ~800°C to form MoOS$_2$ phase.

$$2MoO_2 + 3S \rightarrow 2MoOS_2 + SO_2 \uparrow \tag{2}$$

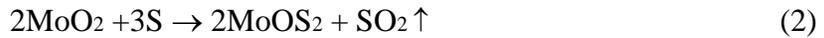

And finally, MoOS$_2$ crystals further sulfurize layer-by-layer at ~900°C to obtain 2D-MoS$_2$ crystals:

$$2MoOS_2 + S \rightarrow 2MoS_2 + SO_2 \uparrow \tag{3}$$

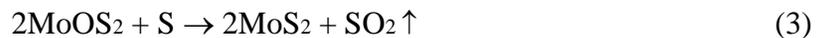

## 3.2. Growth of MoS$_2$ using S and AHM

Tao *et al.*[21] suggested that Mo vapor precursors, generated from AHM [(NH$_4$)$_6$Mo$_7$O$_{24}\cdot$ 4H$_2$O)] instead of MoO$_3$, in low diffusion rate enable growth of low nuclei density which produce large-sized crystalline MoS$_2$ film. So, in this case, MoO$_3$ powder was replaced by AHM. Amount of S precursor and Ar flow were kept fixed at 100 mg and 60 ml/min, respectively, whereas amount of AHM precursor and growth temperature were varied systematically. Initially, at AHM= 20 mg (S to AHM particles ratio ~193:1) and T= 900°C, round shape MoS$_2$ oxy-flakes (~1-10 μm) having a nucleation centre at mid of the flakes, were synthesised (**Fig. 3a**). Furthermore, to convert the oxy-phase round shape MoS$_2$ to triangular MoS$_2$ flakes, amount of AHM was decreased to 15 mg (S to AHM particles ratio ~257:1), and growth temperature increased to 950°C. High-quality large-sized 1L-MoS$_2$ flakes (sky blue triangles, ~1-50 μm) along with bulk MoS$_2$ (bright triangles, ~1-20 μm) (**Fig. 3b**) were observed. When growth temperature was further increased to 1000°C, uniform growth of high-quality 1L- and few bulk MoS$_2$ triangles (~20-80 μm) over a large area of the substrate (~5 mm × 10 mm) were obtained (**Fig. 3c**). **Fig. 3b,c** correspond to high-quality MoS$_2$ flakes grown over a large area obtained at nearly same growth conditions, which implicate high repeatability of the method in the temperature range 950-1000°C. **Fig. 3d-f** correspond to FESEM images of MoOS$_2$ with few MoS$_2$ flakes, bulk and 1L-MoS$_2$ triangular flakes, and large-sized crystalline 1L-MoS$_2$ flakes, respectively. At the high temperature (900-1000°C), pyrolysis occurs that causes molecular breakdown of AHM. Possible pyrolysis reaction of AHM is as follows [21]:

$$(NH_4)_6Mo_7O_{24}\cdot 4H_2O \rightarrow 7MoO_3 + 6NH_3 \uparrow + 7H_2O \uparrow \qquad (4)$$

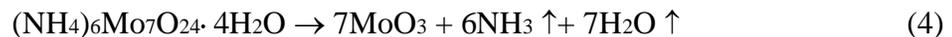

MoO$_3$ vapor particle reacts with S vapors while ammonia gas (NH$_3$) and water vapors (H$_2$O) are flushed out. A very high number of S atoms supply (S to AHM particles ratio ~257:1) was required to synthesise MoS$_2$, so that S atoms can easily react with MoO$_3$ molecules obtained from breakdown of AHM. Further, growth mechanism could be considered similar to MoO$_3$ which produce MoS$_2$ in three aforementioned steps.

### 3.3. Tellurium assisted growth of MoS$_2$

For large scale industrial synthesis of MoS$_2$, there is a pressing need to lower down growth temperature. Gong *et al*.[32] reduced the growth temperature of MoS$_2$ by 200°C by using Te as growth intermediate in two zone CVD furnace. In our work, CVD synthesis of Te-assisted MoS$_2$ crystalline triangular flakes on SiO$_2$/Si substrate at atmospheric pressure were optimised at a constant loading of source precursors at Te= 15 mg, S= 100 mg, and MoO$_3$= 15 mg (S to MoO$_3$ to Te particles ratio ~30:1:1.13) with different growth temperature. A mixture of MoO$_3$ and Te powder was loaded in quartz tube underneath the SiO$_2$/Si substrate (face-down manner). At a temperature of 650°C, triangular star shape bulk MoS$_2$ flakes (bright blue triangles, ~10-30 μm) were grown over the island of 1L-MoS$_2$ flakes (purple triangles, ~20-70 μm) (**Fig. 4a**). **Fig. 4b** shows the OM image of crystalline 1L-MoS$_2$ synthesised at 700°C with lateral size up to 50 μm. **Fig. 4c** represents very large size (~50-150 μm) highly crystalline uniformly grown pine shape triangular bulk MoS$_2$ flakes over the large area (~5 mm × 6 mm). Furthermore, **Fig. 4d-f** are FESEM images corresponding to **Fig. 4a-c**. Initially, at temperature lower than 650°C, small fraction of MoO$_3$ powder dissolve in molten Te which make unstable ternary alloy MoS$_x$Te$_{2-x}$ under continuous flow of S atoms[32,35]. This unstable alloy acts as a reaction intermediate which at higher temperature result in 2D-MoS$_2$ flakes, and byproducts are flushed out by Ar gas[32,34]. Without the use of Te powder, MoS$_2$ growth temperature was 900°C. Hence, introduction of Te accelerates growth and lower down MoS$_2$ growth temperature by ~250°C which could be useful for large scale MoS$_2$ growth[34].

4. **Raman analysis**

Raman spectroscopy was carried out on as-grown MoS$_2$ flakes. There are first-order four Raman-active vibrational modes (E$_{2g}^2$, E$_{1g}$, E$_{2g}^1$, and A$_{1g}$) in MoS$_2$[33]. E$_{2g}^1$ mode corresponds to in-plane vibration of two S atoms in opposite direction to Mo atom, and A$_{1g}$ mode represents out-of-plane vibration of two S atoms in opposite direction while Mo atom remains at rest[33,36]. E$_{2g}^1$ and A$_{1g}$ as fundamental first-order Raman modes of MoS$_2$ were observed while rest of the two modes were undetectable due to the limited rejection of the Rayleigh scattered radiation in our Renishaw-inVia-confocal-Raman microscope-6260.

Initially, we observed the oxy-phase of MoO$_3$ (MoO$_2$) that resulted in many Raman peaks around at 128, 206, 230, 352, 367, 466, 498, 571, 592, and 746 cm$_{-1}$ (**Fig. 5a**). These Raman peaks of MoO$_2$ match well with reported results[37]. Raman spectra of MoO$_2$ was taken from the same sample whose OM and SEM images are shown in **Fig. 2a,d**, respectively. Furthermore, an increased amount of S resulted in the oxy-phase of MoS$_2$ (MoOS$_2$) whose Raman peaks are observed at around 127, 206, 231, 367, 474, and 499 cm$_{-1}$ for MoOS$_2$[25], and 384 cm$_{-1}$ and 409 cm$_{-1}$ for bulk MoS$_2$ (**Fig. 5b**). Raman spectra of MoOS$_2$ was taken from the same sample whose OM and SEM images are shown in **Fig. 2b,e**, respectively. Peak frequency difference (Δk) in these E$_{2g}^1$ and A$_{1g}$ mode is closely related to a number of layers[33,38]. For a given λ, the difference between E$_{2g}^1$ and A$_{1g}$ modes decreases with decreasing thickness of MoS$_2$. It has been reported that Δk of E$_{2g}^1$ and A$_{1g}$ modes for monolayer is < 21 cm$_{-1}$, ~21-22 cm$_{-1}$ for bilayer, and Δk > 25 cm$_{-1}$ for bulk MoS$_2$[32,39]. **Fig. 5c** represents Raman spectra of crystalline 1L, 2L, and bulk MoS$_2$ synthesised using $MoO_3$ and S powder. For 1L-$MoS_2$, $E_{2g}^1$ and $A_{1g}$ Raman modes observed at 384 and 404.2 cm$^{-1}$ and their $\Delta k$ for these modes is 20.2 cm$^{-1}$ which corresponds to monolayer $MoS_2$ (**Fig. 5c**). For 2L-$MoS_2$, $E_{2g}^1$ and $A_{1g}$ mode were observed at 383.3 and 404.5 cm$^{-1}$, and for bulk $MoS_2$, modes are observed at 382.7 and 409.4 cm$^{-1}$, respectively (**Fig. 5c**). As the layer numbers of $MoS_2$ increase, $A_{1g}$ Raman mode shows blue shifts while $E_{2g}^1$ mode shows red shifts. Blue shifts of $A_{1g}$ mode can be attributed to increased restoring force because of interlayer van der Waals interaction between layers of $MoS_2$`. Furthermore, this restoring force consideration does not predict the behaviour of $E_{2g}^1$ mode (in-plane vibration) so, it is assumed that this anomaly is caused by stacking induced structural change and/or long-range coulombic interaction[33]. Greater intensity of $A_{1g}$ mode than $E_{2g}^1$ for bulk $MoS_2$ shows stronger out-of-plane vibration of $MoS_2$ molecule while comparatively equal intensities of $A_{1g}$ and $E_{2g}^1$ modes represents weak coupling between electronic transition at K-point[38] (**Fig. 5c**). Raman spectra of CVD $MoS_2$ synthesised by AHM powder and Te-assisted are shown in **Fig. 5d,e**, respectively. Raman intensity of $A_{1g}$ mode is considerably higher than the $E_{2g}^1$ mode for 1L, 2L, and bulk $MoS_2$ for AHM powder and Te-assisted CVD $MoS_2$ depicting strong out of plane vibration of $MoS_2$ (**Fig. 5d,e**).

5. **Photoluminescence analysis**

Photoluminescence (PL) spectroscopy was carried out with 532 nm laser line on 1L-$MoS_2$ triangular flakes to determine the bandgap. PL spectra of $MoS_2$ triangular flakes grown via a different combination of precursors are shown in **Fig. 5f**. PL spectrum (1), (2), and (3) correspond to $MoS_2$ flakes synthesised by $MoO_3$, AHM, and Te-assisted, respectively. PL spectrum (1) shows A exciton peak at 668 nm (1.856 eV) and B exciton peak around 620 nm (2.0 eV). Spin-orbit coupling causes splitting of valence band which leads to A and B excitonic transitions[40,41]. These two A and B excitons are generated because of radiative recombination of electrons and holes in a neutral exciton at the Brillouin zone K-point[40]. The energy of valence band splitting is 144 meV (2.0-1.856 eV), and A exciton transition occur from top of valence band while B exciton from the bottom of the valence band. PL spectrum (2) and (3) correspond to 1L-MoS$_2$ flakes synthesised by AHM and Te-assisted, respectively. A and B excitons position were observed at 666 nm (1.861 eV) and 622 nm (1.994 eV), respectively for AHM precursor and at 680 nm (1.823 eV) and 634 nm (1.956 eV) for Te-assisted growth. Valence band splitting for both AHM precursor and Te assisted growth was observed at 133 meV.

6. **Morphological study of as-synthesised MoS$_2$ flakes**

Tapping mode atomic force microscopy (AFM) was carried out to see surface morphology and to determine the thickness of MoS$_2$ flakes. **Fig. 6a-c** are AFM height images corresponding to CVD synthesised MoS$_2$ flakes grown by precursors MoO$_3$, AHM, and Te-assisted, respectively. AFM height images are taken from the same samples for which OM images are shown in **Fig. 2c, 3c, and 4c**, and SEM images are shown in **Fig. 2i, 3f and 4f**. AFM height profiles of corresponding AFM images are inserted in respective Figures. In **Fig. 6a**, AFM height image shows the thickness of 1L-MoS$_2$ to be 0.88 nm (with high uniformity of surface) nearly similar to the theoretical value of 0.62 nm[42]. While CVD MoS$_2$ synthesised by AHM and Te-assisted give monolayer thickness to be 1.05 nm and 1.25 nm, respectively (**Fig. 6b,c**). Relatively higher value of the thickness of 1L-MoS$_2$ are regarded as SiO$_2$/Si substrate has some absorbents on its surface[33]. Comparing surface topography of **Fig. 6a-c**, it is observed that triangular flake of **Fig. 6b** has some 2L-MoS$_2$ grown over 1L-MoS$_2$ flake, and corners of the 1L-MoS$_2$ triangle are truncated. Similarly, the triangular 1L-MoS$_2$ flakes are shown in **Fig. 6c** has irregular sides, and bulk MoS$_2$ triangles (bright triangles) of very small size are grown over 1L-MoS$_2$ flake. Analysing the surface roughness, homogeneity, and uniformity of these flakes make it clear that the morphology of MoS$_2$ flakes synthesised by MoO$_3$ precursors is smoother with better uniformity as compared to AHM and Te-assisted growth. AFM height images are taken from the same samples for which optical images are shown in **Fig. 2d,g** and **i**, and SEM images are shown in **Fig. 3f, 4c** and **4f**.

## 4. Conclusions

Effect of different precursors on CVD growth of MoS$_2$ on SiO$_2$/Si substrate was studied by changing growth temperature and precursors ratio. Highly crystalline large-sized 1L-MoS$_2$ flakes were obtained at S to MoO$_3$ particles ratio of ~30:1. With the introduction of Te, MoS$_2$ growth temperature is lowered down by ~250°C that could be useful in low-temperature large scale growth of MoS$_2$ flakes. For large-sized crystalline 1L-MoS$_2$ flakes, MoO$_3$ powder is a better choice, and for large area bulk MoS$_2$, both AHM powder and Te-assisted synthesis both can be chosen. MoS$_2$ flakes grown by MoO$_3$ and S precursors shows better uniformity, crystallinity, and reproducibility as compared to AHM and Te-assisted growth. Photoluminescence reveals that the energy of valence band splitting (144 meV) is higher for MoO$_3$ than AHM and Te-assisted 1L-MoS$_2$ (133 meV) synthesis. With the help of Raman spectroscopy, we propose three-step chemical reaction of evolution of MoS$_2$ from vapor phase MoO$_3$ micro-particles. This study can be further used to fabricate MoS$_2$ based high-performance electronic devices such as sensors, thin film transistors, and photodetectors.

**Acknowledgements**

The authors gratefully acknowledge the Nanoscale Research Facility (NRF) and FIST-UFO scheme at the Indian Institute of Technology Delhi for providing AFM characterizations, and Raman and photoluminescence measurements, respectively. Aditya Singh would like to thank the University Grants Commission (UGC), India for providing research fellowship.

[41] D. Kaplan, Y. Gong, K. Mills, V. Swaminathan, P.M. Ajayan, S. Shirodkar, E. Kaxiras, Excitation intensity dependence of photoluminescence from monolayers of $MoS_2$ and $WS_2$/$MoS_2$ heterostructures, 2D Mater. 3 (2016) 015005. doi:10.1088/2053-1583/3/1/015005.

[42] R.F. Frindt, Single Crystals of $MoS_2$ Several Molecular Layers Thick, J. Appl. Phys. 37 (1966) 1928–1929. doi:10.1063/1.1708627.

**Figures**

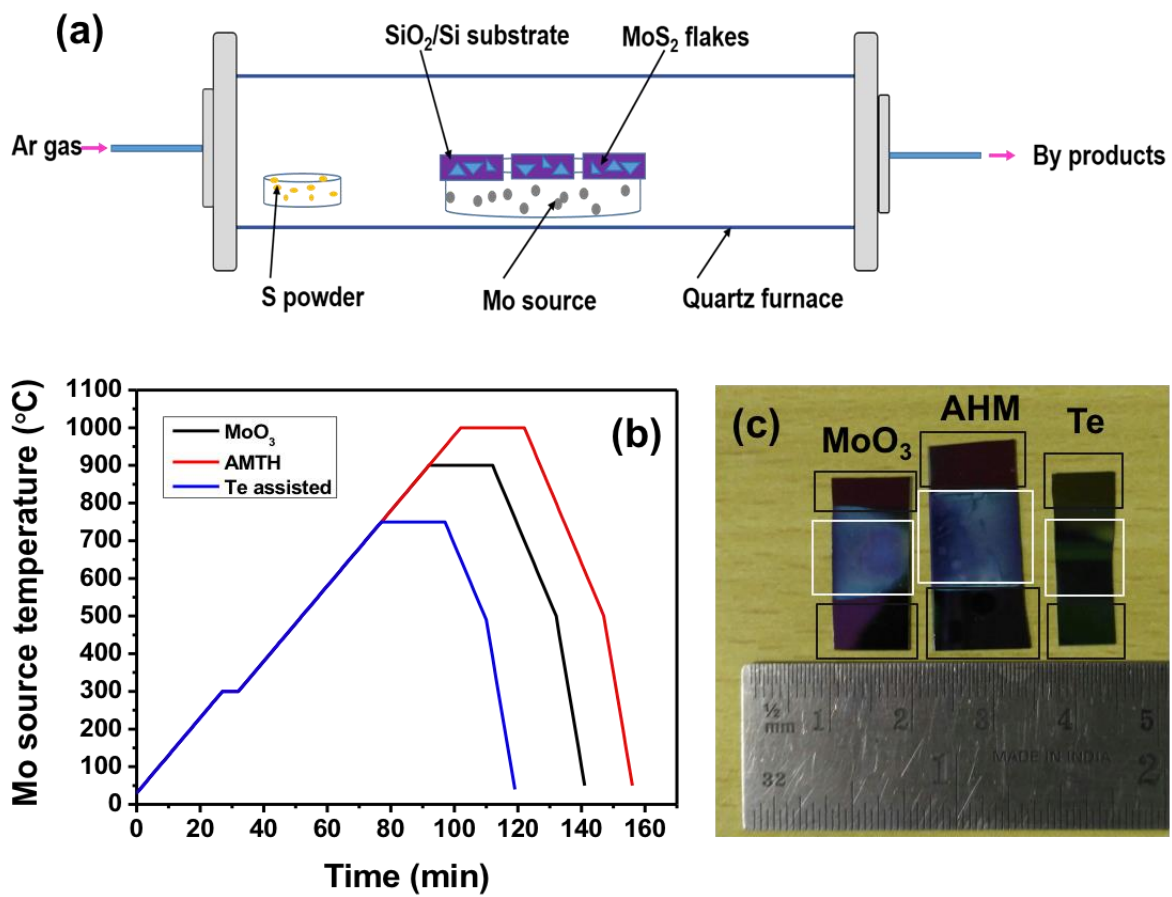

**Fig. 1.** (a) Schematic illustration of the inside view of the quartz tube of single zone mini CVD setup. (b) CVD temperature-time programme graph for different Mo precursors. (c) Image of as-grown $MoS_2$ samples on $SiO_2$/Si substrate showing large area growth. Black boxes represent regions of as-grown $MoS_2$ flakes while white boxes show $MoO_2$ and $MoOS_2$ grown regions

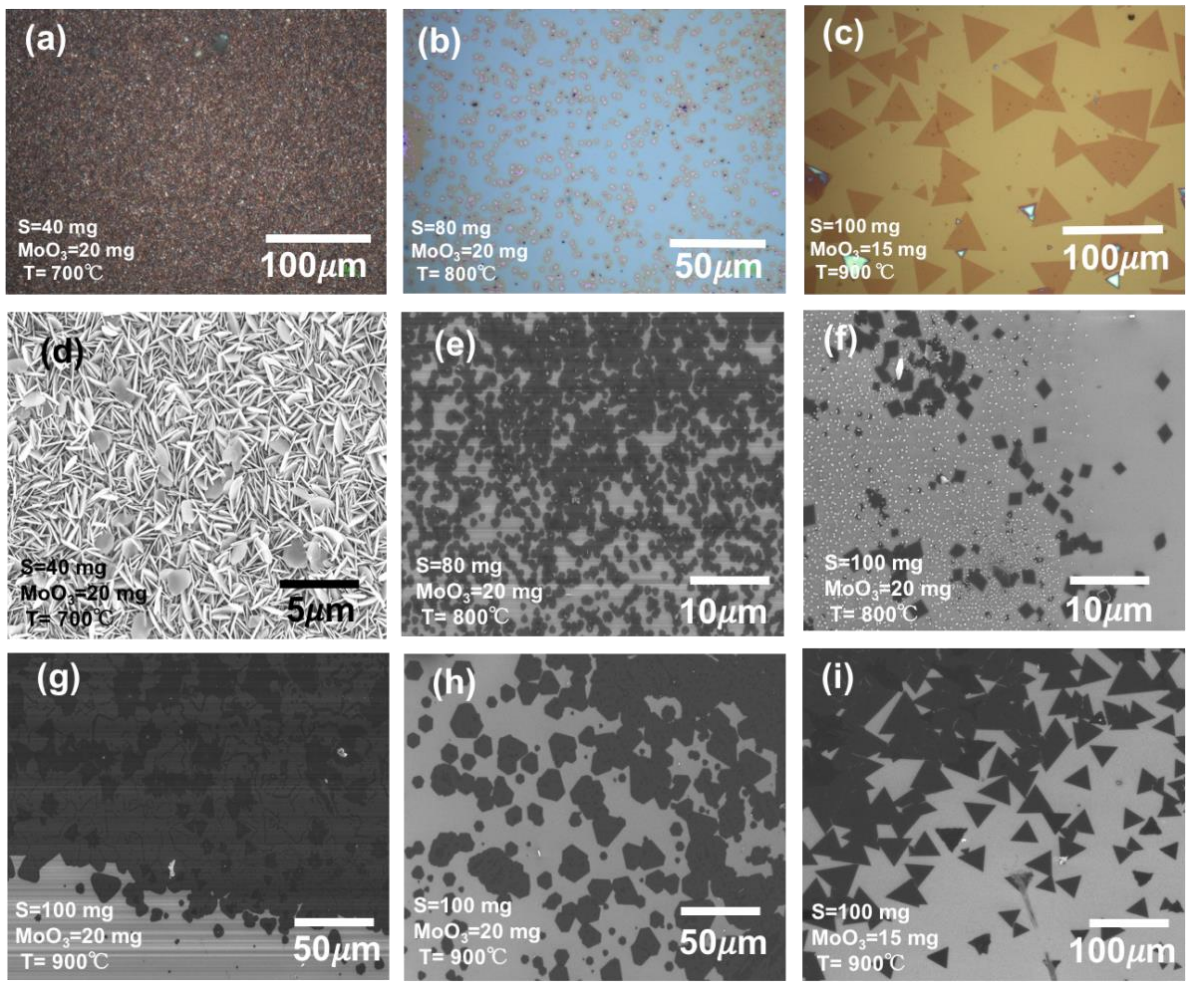

**Fig. 2.** Optical microscopy (OM) and scanning electron microscopy (SEM) images of different MoS$_2$ phase synthesised by MoO$_3$ as Mo source. (a) OM image of as-grown oxy-phase MoO$_2$, (b) MoOS$_2$, and (c) 1L-MoS$_2$, depicting the effect of precursors ratio and growth temperature on MoO$_3$. (d)-(i) SEM images of the evolution of MoS$_2$ from MoO$_2$ to highly crystalline 1L-MoS$_2$ flakes. Images (g) and (h) are captured at different locations of the same sample

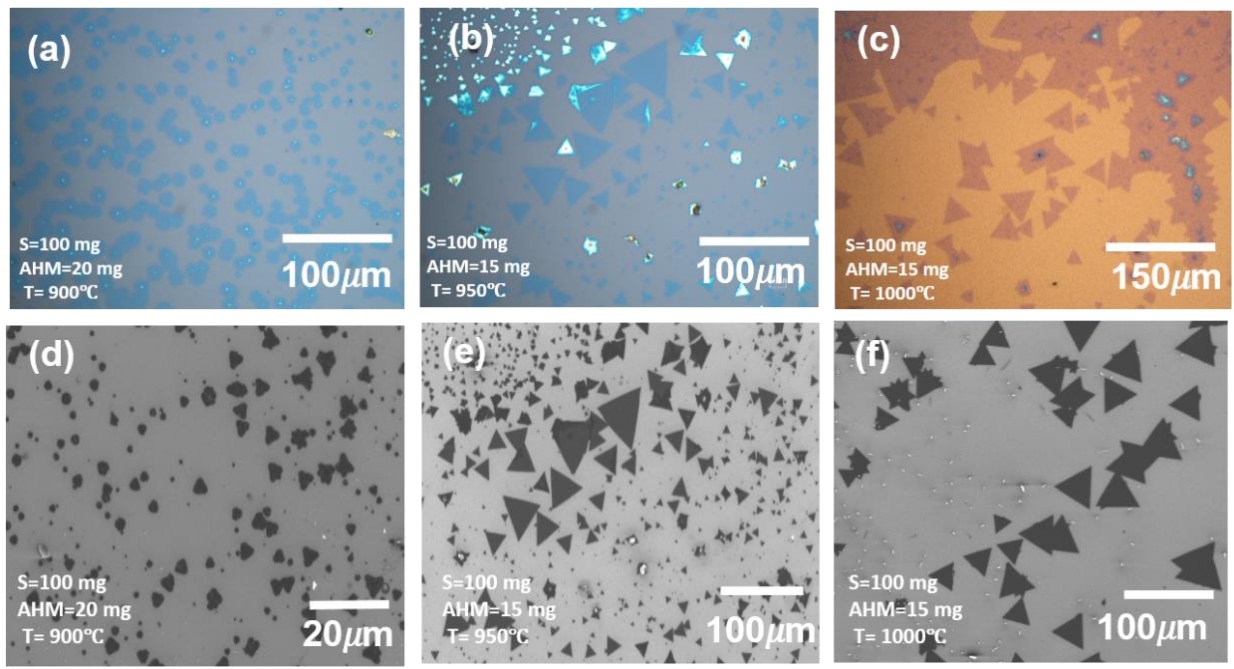

**Fig. 3.** Optical and SEM images representing different MoS$_2$ phases synthesised by AHM precursor as Mo source. (a) OM image of round shape oxy-phase MoS$_2$ micro-crystals acting as a nucleation centres for growth. (b) Bulk and 1L-MoS$_2$ flakes. (c) Large-sized crystalline uniform 1L-MoS$_2$ flakes. (d)-(f) SEM images corresponding to OM images showing the evolution of MoS$_2$ from MoO$_2$ to highly crystalline monolayer synthesised by AHM precursor as Mo source

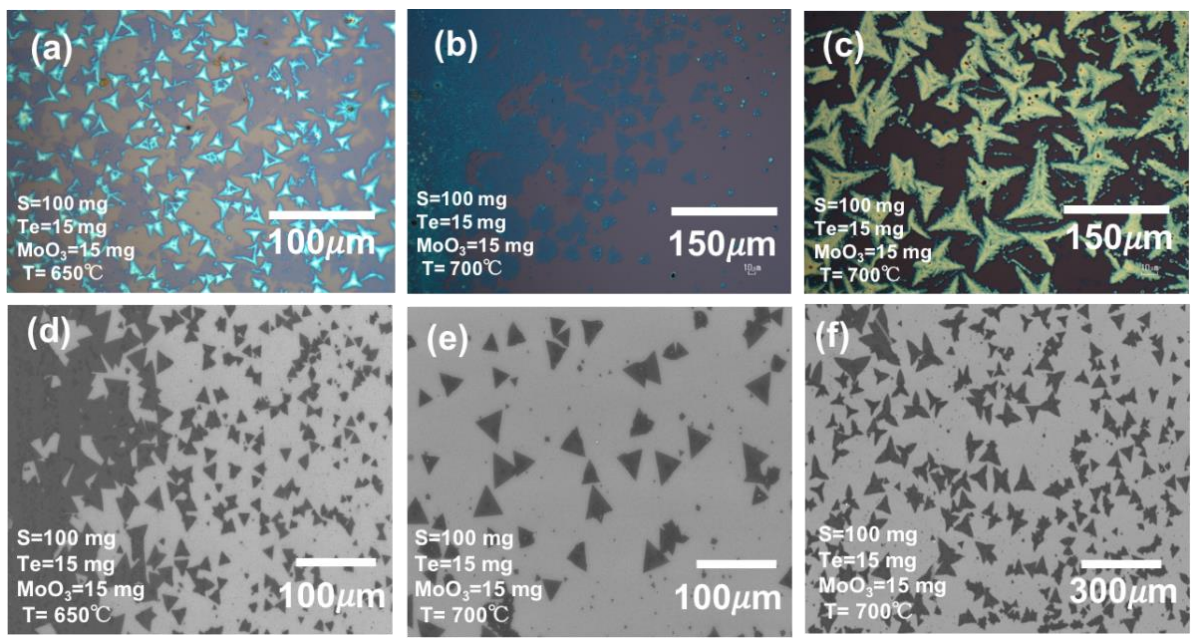

**Fig. 4.** Optical and SEM images of Te-assisted MoS$_2$ flakes. (a)-(c) optical micrographs and (d)-(f) are their corresponding SEM images of highly crystalline Te-assisted MoS$_2$ flakes

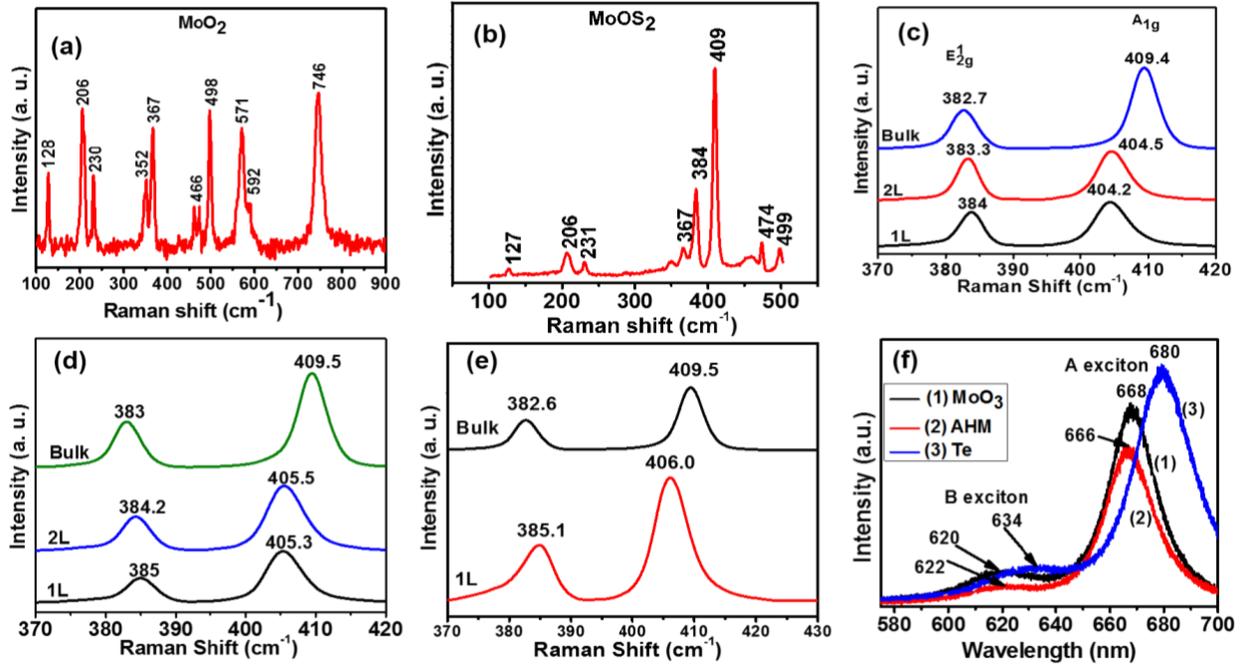

**Fig. 5.** Raman and photoluminescence measurements characterised by 532 nm laser. (a) and (b) Raman spectra of micro particles of $MoO_2$ and $MoOS_2$, respectively. (c) Raman spectrum of crystalline 1L, 2L, and bulk $MoS_2$ synthesised using $MoO_3$ precursor at 900°C. $E^1_{2g}$ and $A_{1g}$ Raman modes observed at 384 and 404.2 cm$^{-1}$ ($\Delta k = 20.2$ cm$^{-1}$). (d) and (e) Raman spectra of CVD $MoS_2$ synthesised by AHM powder and Te-assisted growth, respectively. (f) PL spectrum of $MoS_2$ triangular $MoS_2$ flakes grown via a different combination of precursors. PL spectrum (1), (2), and (3) correspond to $MoS_2$ synthesised by $MoO_3$, AHM and Te assisted, respectively. In all these three PL spectrums, higher and lower energy peak corresponds to B exciton and A exciton, respectively

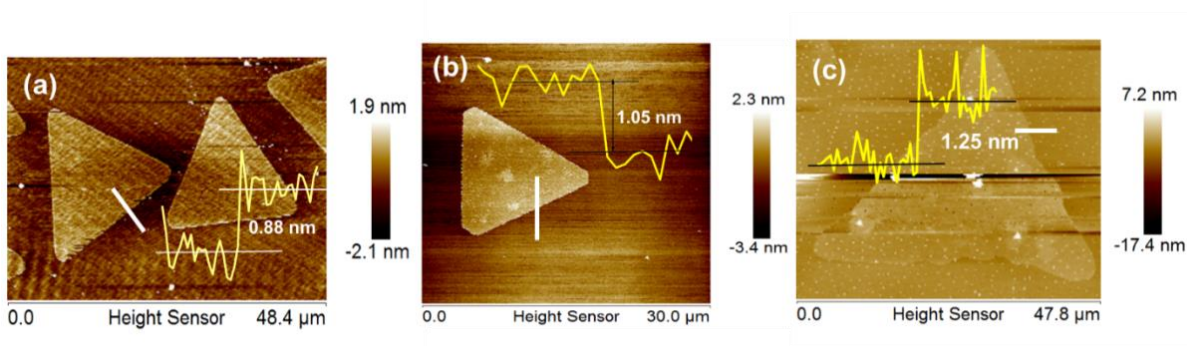

**Fig. 6.** AFM images and height profiles of as-grown MoS$_2$ flakes. AFM height profiles of corresponding AFM images are inserted in respective Fig. and white line in AFM images show the path of height profile measurements. (a) triangular crystalline, uniform 1L-MoS$_2$ synthesised using MoO$_3$ having a layer thickness of 0.88 nm. (b) crystalline truncated 1L-MoS$_2$ triangle of 1.05 nm grown by AHM. (c) Te-assisted pine shape triangular flake of 1L-MoS$_2$ (thickness ~1.25 nm) grown at lower temperature (750°C)